\documentclass[10pt]{iopart}

\usepackage{iopams}  
\usepackage{graphicx}
\usepackage{amssymb}
\usepackage{epstopdf}
\usepackage{float}
\usepackage{color}
\usepackage{url}
\expandafter\let\csname equation*\endcsname\relax
\expandafter\let\csname endequation*\endcsname\relax
\ioptwocol
\usepackage{amsmath}

\newcommand{\be}{\begin{equation}}
\newcommand{\ee}{\end{equation}}

\newcommand{\bra}[1]{\langle{#1}|}
\newcommand{\ket}[1]{|{#1}\rangle}

\newcommand{\E}[1][\empty]{
  \ifthenelse{\equal{#1}{\empty}}
    {\mathbb{E}}
    {\mathbb{E}\left( #1 \right)}
}

\renewcommand{\exp}[1][\empty]{
  \ifthenelse{\equal{#1}{\empty}}
    {\mathrm{exp}}
    {\mathrm{e}^{#1}}
}

\newcommand{\psit}[1][\empty]{%
  \ifthenelse{\equal{#1}{\empty}}
    {\psi_t}
    {\psi_t^{(#1)}}
}

\newcommand{\npsit}[1][\empty]{%
  \ifthenelse{\equal{#1}{\empty}}
    {\tilde\psi_t}
    {\tilde\psi_t^{(#1)}}
}

\newcommand{\up}{\uparrow}
\newcommand{\down}{\downarrow}
\usepackage{ulem}  

\newcommand{\oalex}[1]{{\color{blue}{}}}

\definecolor{olive}{RGB}{107,142,35}

\definecolor{orange}{RGB}{255,139,61}

\usepackage{cite}

\begin{document}
\title{Delocalization in two and three-dimensional Rydberg gases}
\author{G. Abumwis, Matthew T. Eiles, Alexander Eisfeld*}
\address{Max-Planck-Institut f\"ur Physik komplexer Systeme, N\"othnitzer Str.\ 38,
D-01187 Dresden, Germany }

\ead{\mailto{eisfeld@mpg.pks.de*}}

\date{\today}
\begin{abstract} 
As was recently shown in Ref. \cite{ourPRL}, many eigenstates of a random Rydberg gas with resonant dipole-dipole interactions are highly delocalized.  Although the high degree of delocalization is generic to various types of power-law interactions and to both two and three-dimensional systems, in their detailed aspects the coherence distributions are sensitive to these parameters and vary dramatically between different systems.  We calculate the eigenstates of both two and three-dimensional gases and quantify their delocalization throughout the atoms in the gas using a coherence measure. By contrasting the angular dependence of the dipole-dipole interaction with an isotropic interaction we obtain additional information about the generic physical principles underlying random interacting systems. We also investigate the density of states and microwave absorption spectra to obtain information about the types of measurements where these delocalized states play a role, and to check that these delocalized eigenstates are robust against various types of perturbation.
\end{abstract}

\section{Introduction}
Resonant dipolar interactions are ubiquitous in nature. They are the subject of extensive study in molecular aggregates \cite{SaEiVa13_21_} and photosynthetic systems \cite{AmVaGr00__}. 
These systems are typically quite regularly arranged and exhibit excitonic states that are delocalized over several particles.
Such regular arrangements have also been studied in other systems, for example in ultracold molecules \cite{yan2013} and Rydberg atoms \cite{BaLaRa15_113002_}.
A renewed interest in \textit{random} arrangements has developed in recent years, stemming from measurements on very dilute gases containing both ground state atoms and atoms in their first excited state  \cite{DaRiLi12_193201_,BrBiSt15_053412_,BrEiBa19_-_,Cundiff2016,yu2019long,Mu16_041102_,LiBrSt17_052509_}. A paradigmatic interacting random system is the single-exciton scenario, where one $\ket\up$ excitation is added to an assembly of $N-1$ particles all in the state $\ket\down$. In the absence of interactions, the single-exciton wave functions are of the form 
\begin{equation}
\ket{n}=\ket{\down}_1\cdots\ket{\up}_n\cdots\ket{\down},
\end{equation}
where the particle at position $n$ is promoted to the $\up$ state.  The single-exciton Hamiltonian takes the form
\begin{align}
\label{eq:ham}
H&= \sum_{n=1}^N \epsilon_n \ket{n}\bra{n}+\sum_n \sum_{m\ne n} V_{nm}(\vec{R}_n,\vec{R}_m)\ket{n}\bra{m}.
\end{align}
The excitation is shared between sites as described by the collective eigenstate
\begin{equation}
\ket{\psi_\ell}=\sum_n c_n^{(\ell)} \ket{n}.
\end{equation} 

An ultracold gas of Rydberg atoms is an ideal system to study these strongly-interacting coherent systems \cite{Pillet1998,GallagherFroze1998,Mourachko2004,scholak2014spectral}. The long-range nature of Rydberg interactions and the sufficiently long atomic lifetimes allow coherent processes to take place on experimental timescales without any significant decoherence due to thermal motion or spontaneous decay \cite{Pilleta,Singer2005,Weber_2017,MartinDDSpec,Park2011}.  The interaction between two well-separated Rydberg atoms is determined by the multipole expansion of each Rydberg charge cloud \cite{Weber_2017}. Due to the large interatomic separations at typical densities, these interactions are usually small compared to the energy separation between states with different principal quantum number. This implies that an essential state picture can  be adopted. In our case, the Hilbert space of each atom is restricted to only include two Rydberg states coupled by the dipole matrix element.  For concreteness, we consider the $s$ and $p$ Rydberg states with the same $\nu$. We label these states $\ket\up$ and $\ket\down$, respectively. This two-state approximation is adequate for most parameter regimes, but it can be invalidated in special circumstances, such as at very high density or principal quantum number, in the presence of significant external fields, or if unfavorable parameters are chosen, leading to accidental near-degeneracies with states outside of our two-state subspace \cite{Robicheaux2004}.

Recently, we studied the eigenstates of a three-dimensional random gas of $N$ Rydberg atoms and focused in particular on their delocalization as described by the distribution of coefficients $c_n^{(\ell)}$ \cite{ourPRL}. We observed that the majority of these collective states are remarkably delocalized, having coherences --  a measure of delocalization -- which grow linearly with the number of gas particles. This delocalization in spite of the random arrangement of particles is facilitated by the dipole-dipole interactions. The random spatial arrangement of the atoms lead to nearby clusters of two or more atoms. These interact strongly due to their proximity and energetically decouple from the remaining system, which now has no clusters and resembles a network with quasi-homogeneous spacing. Because of the long-range nature of the dipole-dipole interaction the atoms remaining in this network have non-negligible interactions even at these larger separations, and can therefore delocalize over many sites. 

While Ref.~\cite{ourPRL} focuses on the experimentally most relevant scenario of anisotropic dipole-dipole interactions in a \textit{three-dimensional} (3D) gas, in the present work we additionally study two-dimensional (2D) arrangements and isotropic interactions.  A two-dimensional gas can be realized in present-day Rydberg experiments with sufficiently tight confinement \cite{chomaz2015emergence}, and provides a very useful experimental complement to the three-dimensional gas as it allows one to modify the isotropy of the interaction in a simple way by varying the relative angle between an applied magnetic field and the confinement axis. The 2D case also relates to the investigation of molecules randomly placed on surfaces \cite{MPaEi13_044302_,MPaMa13_064703_}. Much additional information about the nature of the delocalized states observed in Ref. \cite{ourPRL} can be gleaned from the study of other power-law interactions, system dimensions, and interaction isotropy, as we undertake in the present work.  We also study the dependence of the localization on the structure of the interaction Hamiltonian, which can be modified due to disorder or the Rydberg blockade. These effects can either reduce or increase the degree of localization. Through these various manipulations of the system's Hamiltonian we can build a body of observations which can reveal trends and dependencies of the localization in the Rydberg gas and embed our conclusions in the broader context of strongly interacting finite systems.   

The paper is organized as follows. In Section \ref{sec:theory} we outline the theoretical background for this system, describing the interactions, absorption spectrum, various measures to quantify the system delocalization, and the details of the numerical simulation. This is followed in Sec. \ref{sec:results} by a characterization of the eigenstates obtained by this calculation in terms of their coherence distributions and robustness to disorder. Sec. \ref{sec:dos} compares the density of states and absorption distributions for the various scenarios, and Sec. \ref{sec:power} studies the dependence of the eigenstate properties on the nature of the long-range interactions. In Sec. \ref{sec:analyt} we consider some simplistic analytic models of this system in order to qualitatively explain some of the observed behavior. We conclude in Sec. \ref{sec:conc}.

\section{Theoretical background}
\label{sec:theory}
\subsection{Long-range interactions in the essential state Hamiltonian}
\label{sec:eqns}
The form of the $s-p$ Rydberg interaction is
\begin{equation}
\label{eq:dipdip}
V_{nm}(\vec{R}_n,\vec{R}_m)= \frac{1}{3}\frac {\mu^2}{\big|\vec R_n-\vec R_m\big|^\alpha}\, f(\theta_{nm}),
\end{equation}
where $\mu$ is the transition dipole between $\up$ and $\down$ states and $\alpha=3$ since this is a resonant dipole interaction. The interaction is anisotropic, $f(\theta_{nm})=(1-3\cos^2\theta_{nm})$. In general, the presence of degenerate magnetic sublevels in the $\down$ and $\up$ states result in this operator taking on a tensorial form \cite{PaTaCl11_22704_,Robicheaux2004}; to simplify it we have assumed that a magnetic field of a few tens of Gauss is applied to the gas. 
This field strength is not strong enough to modify the Rydberg states themselves, but shifts each $m_l$ level at a rate of $1.4$MHz/G via the Zeeman term  \cite{BaLaRa15_113002_}.  In this fashion, one can excite only the $m_l = 0$ states and thus simplify the angular dependence of the interaction to a function $f(\theta_{nm})/3$, depending only on the relative angle between the interparticle distance and the magnetic field. The factor $\frac{1}{3}$ comes also from this separation of degenerate $m$-sublevels.  In the present work we study also isotropic interactions, having $f(\theta_{nm}) = 1$, as can be implemented experimentally in the two-dimensional gas by orienting the magnetic field perpendicular to the plane of atoms. 
Since the key properties of Rydberg atoms scale with the level of excitation and can thereby be tuned extensively, experimentalists can access a wide parameter range \cite{low2012experimental,saffmanJPB,Saffman}. 
We work in a parameter regime where motion induced by the dipole-dipole interaction can be neglected on the time scales of interest.
Note that strong forces can be present between atoms in close proximity, leading to acceleration and a breakdown of the frozen gas approximation \cite{LiTaJa06_27_,Li2006,Plasma1,Plasma2, AtEiRo08_045030_}. For extended atomic networks, conical intersections might play a role \cite{WEiRo11_153002_}.

\subsection{Absorption spectrum}

For a given random realization of atomic positions, $H$ is diagonalized to obtain the eigenstates. The transition frequencies between $s$ and $p$ states range from a few GHz down to hundreds of MHz. Therefore, the eigenstates can be excited from the initial state, $\ket{G}= \ket{\down}_1\dots\ket{\down_N}$, by   microwave absorption. A specific state $\ket{\psi_\ell}$ can be excited provided that this microwave frequency is detuned from the atomic transition by that state's eigenenergy and that the transition strength, $\mathcal{A}_\ell$, is finite. For a microwave where the electric field component is aligned parallel to the external magnetic field, the transition strength is 
\begin{equation}
\mathcal{A}_\ell = \left|\bra{\psi_\ell}\sum c_{n}^{(\ell)}\ket{G}\right|^2= \Big|\sum_n c_{n}^{(\ell)}\Big|^2.
\label{eq:absorp}
\end{equation}
This assumes identical transition dipoles and weak interactions with the microwave.
We consider a situation where the extent of the Rydberg gas is small compared to the wavelength of the microwave radiation so that phase variations of the field can be neglected. 
  The presence of large absorption strengths requires a large delocalization: an absorption of $N$ can only be achieved by a perfectly delocalized collection of atoms, i.e. $c_{n\ell}=N^{-1/2}$. However, the reverse is not true: a large delocalization does not imply a large absorption, and hence the absorption spectrum is not a sensitive probe of delocalized states. 

\subsection{Localization measures}
 The delocalization of a quantum state can be quantified using a variety of measures. The two most common ones, due to their ease of interpretation, are the coherence and the participation ratio. Using the density matrix, {$\rho^{(\ell)}=\ket{\psi^{(\ell)}}\bra{\psi^{(\ell)}}$}, the coherence is defined as \cite{BaCrPl14_140401_}
\begin{equation}
\label{eq:coherence}
\mathcal{C_\ell}=\sum_{n}\sum_{m\ne n} |\rho^{(\ell)}_{nm}|.
\end{equation}
{where $\rho^{(\ell)}_{nm}=c_n^{(\ell)} c_m^{(\ell)} $. 
Note that for our symmetric Hamiltonian the wavefunction coefficients $c_n^{(\ell)}$ can be chosen to be real.}
Using only the wave function itself, the ``participation ratio'' (PR), sometimes called the inverse of the ``inverse paticipation ratio'' (IPR), is
\begin{equation}
N^\mathrm{PR}_\ell=\left[\sum_n \left(\rho_{nn}^{(\ell)}\right)^2. \right]^{-1},
\end{equation}
In general, the value of either measure corresponds roughly to the number of atoms coherently sharing the excitation. The PR for a perfectly delocalized state having equal amplitude on each atom equals $N$, while the coherence is $N-1$. A dimer state, with only two atoms participating, has a coherence of $1$ and an PR of $2$. We show in the appendix that both coherence and PR measures give similar results, but the coherence highlights asymmetries in the distributions and spectra that are less clear in the PR.

\subsection{Numerical implementation of the Rydberg gas }
\label{sec:model}
The positions of the Rydberg atoms are determined by the random positions of the ground state atoms and on the details of the excitation process.  Typically, only a fraction ($\sim 1$\%) of the ground state atoms are selected at random to become Rydberg atoms  \cite{low2012experimental}. By varying the laser power, higher or lower densities can be achieved. 

However, during this process the so-called Rydberg blockade mechanism can influence the spatial distribution of Rydberg atoms \cite{Lukin2001,tong2004local,gaetan2009observation,urban2009observation}. Due to the van der Waals interaction, the interaction energy of two adjacent Rydberg atoms in the $\down$ state can exceed the laser bandwidth and thus prevent their mutual excitation.  The relevant length scale is given by the blockade radius $R_\mathrm{B}$, which scales as $R_\mathrm{B}\propto (\nu^{11}/\Omega_\mathrm{Laser})^{1/6}$. Thus, for a given density of ground state atoms, both the principle quantum number $\nu$ and the laser bandwidth $\Omega_\mathrm{Laser}$ can be used to prevent the excitation of closely-spaced Rydberg atoms. This both reduces the density of $\down$-state atoms relative to the initial ground-state gas density and also removes pronounced inhomogeneities in the gas formed by random clustering \footnote{One can even use this blockade mechanism in conjunction with appropriate laser detuning to form anti-blockaded atoms in regular arrangements \cite{ates2007,pohl2010,schauss2012}}. In the present work we use a simplified description of the blockade mechanism and assume that, when two or more ground state atoms are closer than $R_\mathrm{B}$, only one of them can be excited into a Rydberg state and the others play no role.  This is implemented in our simulation by removing from the initially seeded atoms one from every pair which have a separation smaller than $R_\mathrm{B}$. We therefore neglect the many-body nature of the blockaded state.

In our simulation, the positions of Rydberg atoms are drawn at random from a uniform distribution.  To take into account fluctuations due to the random gas realizations, we average over 2000 total iterations for each calculation. 
We take as the system volume a rectangle of volume $L^D$, where $D=2,3$ is the dimension and $L$ is the side length. For all cases we fix the number of atoms to $N=1000$; for the atomic densities, excitation efficiencies, and dimensions of a typical ultracold gas experiment, this number can typically be produced per shot. As a result, these edge effects crudely model the physical boundaries of the laboratory gas. For certain trapping configurations the distribution of Rydberg atoms can have sharp boundaries within a specifically shaped area \cite{chomaz2015emergence}; on the other hand, in atomic vapours the boundaries can be very diffuse. We have verified that the boundary effects are only minor by reproducing these results with a spherical distribution of atoms and with atoms distributed according to a normal distribution. 

\begin{figure}
\includegraphics[width=\columnwidth]{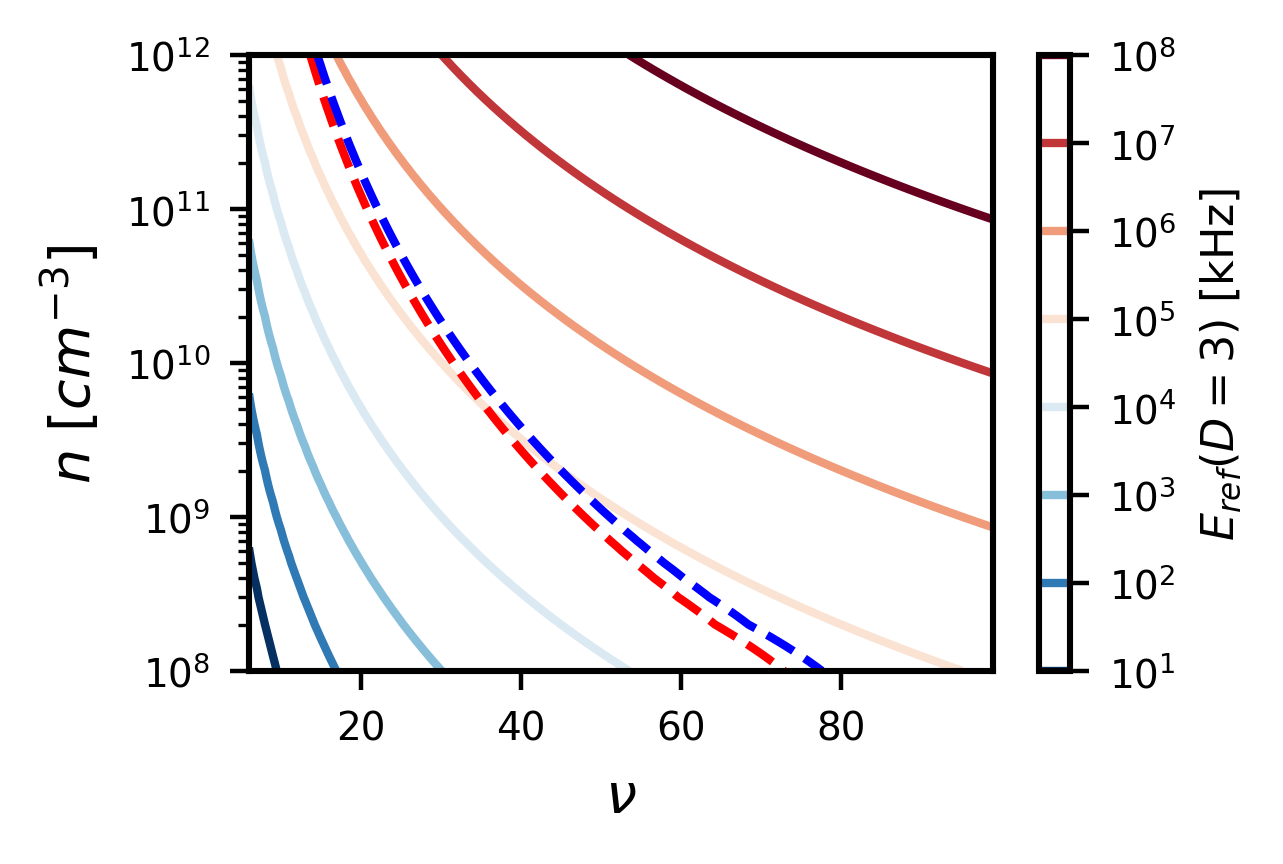}
\caption{The  parameter space for a three-dimensional Rydberg gas of  $^{87}$Rb atoms. Contours of  constant $E_{\rm ref}(3)$ (Eq. \ref{eq:energyunit}) are shown as a function of principal quantum number $\nu$ and Rydberg density $n$. The dashed contours give the parameter range required to achieve a blockade radius $R_B = 0.5$, in units of $a_\text{ref}(3)$, for two different values of $\Omega_\text{Laser}$, $0.5$ MHz (red) and $1$ MHz  (blue). }
\label{fig:parameters}
\end{figure}

The Rydberg  density $n_\text{Ryd}$ is determined by the dimensions of the simulation volume following $n_\text{Ryd}=N/(2R)^D$. We define scaled length and energy units which remove the dependence of the Hamiltonian on density and let us deal with dimensionless quantities only. We use the Wigner-Seitz radius, $a_\mathrm{ref}(D)$, as our unit of length. This is defined in two and three dimensions as
\be
a_\mathrm{ref}(2)=\left(\frac{1}{\pi n_\text{Ryd}}\right)^{1/2};\,\,\,a_\mathrm{ref}(3) = \left(\frac{3}{4\pi n_\text{Ryd}}\right)^{1/3}.
\ee
This unit of length, in turn, defines a characteristic energy scale. The nearest neighbour interaction energy (neglecting the anisotropy of the interaction) between two particles separated by the respective Wigner-Seitz radius is
\be
\label{eq:energyunit}
E_{\rm ref}(D) = \frac{1}{3}\frac{\mu^2}{[a_\text{ref}(D)]^\alpha}.
\ee
We measure energies in units of $E_\mathrm{ref}(D)$.  The density, principal quantum number $\nu$, and $E_\mathrm{ref}(D)$ are all related and can be tuned accordingly to match laboratory constraints, as shown for the case $D=\alpha=3$ in Fig.~\ref{fig:parameters}. For the transition dipole we use $\mu=\mu_0\nu^2$, where $\mu_0$ is the transition dipole between the ground and first excited state.

\section{Characterization of the eigenstates}
\label{sec:results}
In this section we show the main results of this numerical computation: the coherence and absorption distributions and the densities of states for a random Rydberg gas interacting via the dipole-dipole interaction ($\alpha=3$). We present the results from the four possible combinations of anisotropic and isotropic interactions within two- and three-dimensional gases together in order to facilitate a phenomenological comparison. 
\subsection{Coherence and eigenenergy distributions}
\label{sec:cohdist}
\begin{figure}[t]
\includegraphics[width=\columnwidth]{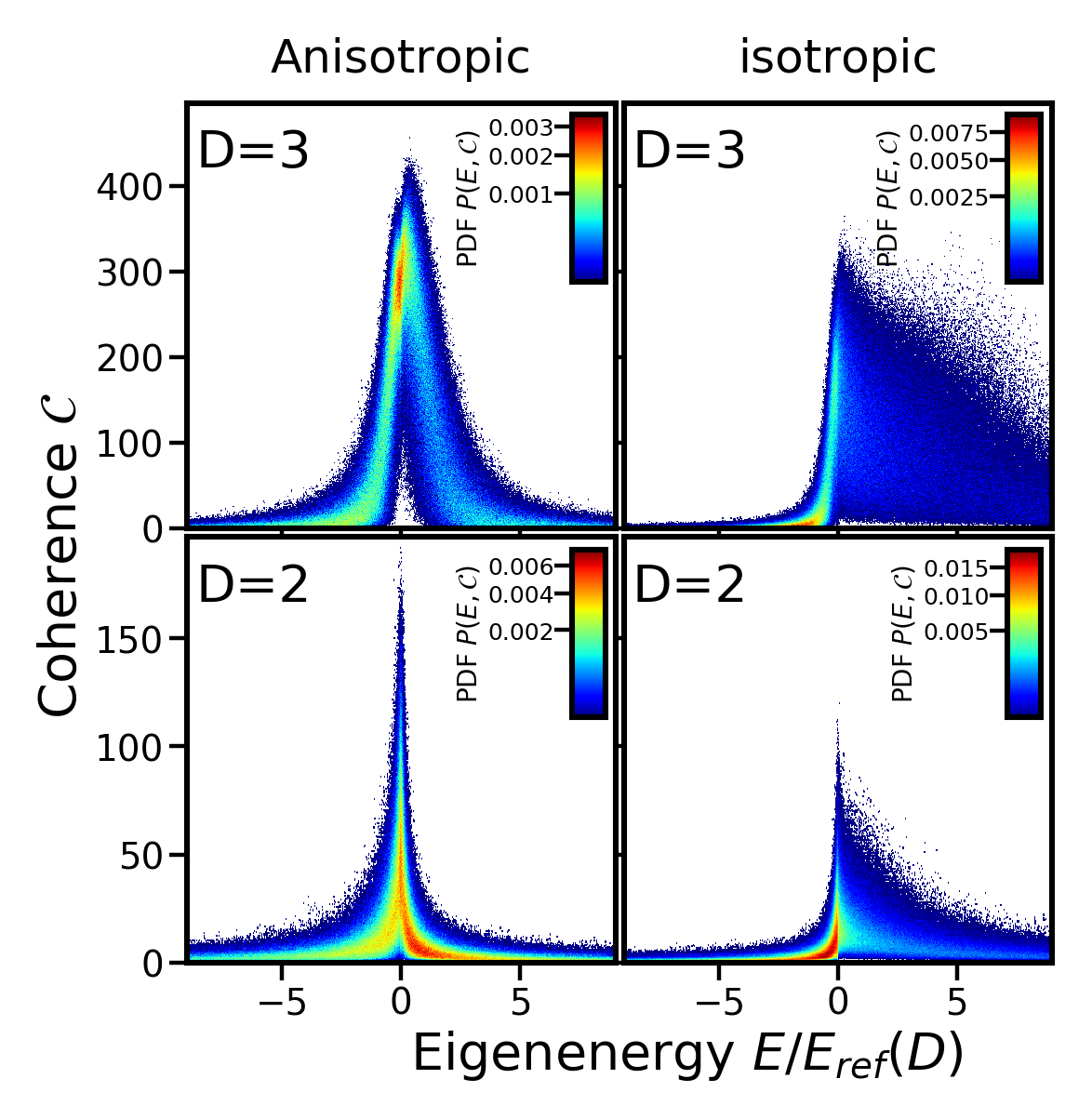}
\caption{Coherence and energy distributions for the $\alpha = 3$ interaction. The coherence binwidth is 0.5 and the energy binwidth is 0.01. The energy unit is $E_\text{ref}(D)$. }
\label{fig:energy}
\end{figure}

Fig.~\ref{fig:energy} displays the distribution of coherences and eigenenergies. The distribution in the top left panel, for the three-dimensional gas with anisotropic interactions, is the main result of Ref.~\cite{ourPRL}. This distribution is narrower on the negative energy side than on the positive, and has a pronounced peak feature at very small negative energies but large coherences. For both positive and negative energies near zero almost all states have a large coherence, and essentially none have a low coherence value. In total, states with coherence values of nearly $N/3$ are remarkably common, and localized states with coherence $<10$ are present in the tails of this distribution. The large energy shifts of these states identify them as small clusters with strong interactions.

The bottom two panels show the results for a two-dimensional gas. For both anisotropic and isotropic interactions the overall coherence lengths are smaller than in three-dimensions, but still exhibit large peaks near zero energy. The isotropic distribution still shows a sharp asymmetry, with a wide spread of delocalized states at positive energies. One notable difference can be seen in the highest probability peaks of both distributions (red color), which are prominent at blue detuning for the anisotropic interaction but at  red detuning for the isotropic interaction. This sign change is related with the fact that the anisotropic case has both positive and negative interactions, while in the isotropic case it is purely non-negative. In section \ref{sec:analyt} we will provide a simple analytical model that provides insight into these findings.

As demonstrated by the top right panel of Fig.~\ref{fig:energy}, this distribution is very sensitive to the anisotropy of the interaction. When the interaction is isotropic the overall delocalization decreases slightly, and the asymmetry about zero energy becomes much stronger. At all negative energies except very close to zero, the states are all highly localized; in contrast, one can find highly delocalized states at all positive energies within the range considered here. Unlike in the anisotropic case, these blue-detuned states are spread almost uniformly over a large range of coherence values, lacking only population in the tiny clusters (very small coherence) or extremely large coherence range.

\begin{figure}[t]
\includegraphics[width=\columnwidth]{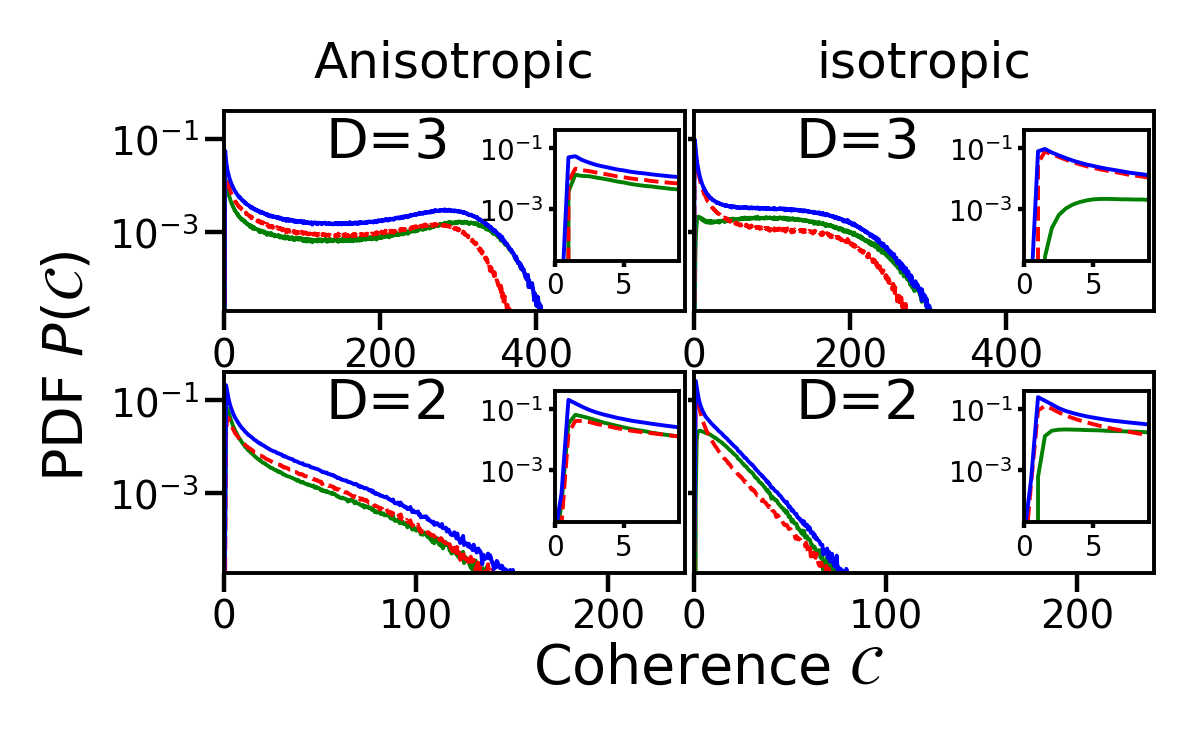}
\caption{\label{fig:CohPosNeg} Marginal distributions of the distributions of Fig.~\ref{fig:energy}, showing the coherence distribution, for three energy ranges. The  curves show the distribution of coherences taking states with all energy (blue), only positive energy (red), and only negative energy (green) into account. 
}
\label{fig:posneg}
\end{figure}

Although the distributions of Fig.~\ref{fig:energy} are asymmetric in several important ways, the marginal coherence distributions integrated over either all negative or all positive energies are very similar (see red and green curves of Fig.~\ref{fig:CohPosNeg}), both resembling the overall marginal distribution plotted as the blue curve in  Fig.~\ref{fig:CohPosNeg}. This distribution has been discussed in detail in Ref. \cite{ourPRL}; it consists of two contributions. The first, highlighted in the inset, is composed of the tightly clustered states with very large interaction strengths. The second is the broad-tailed distribution of delocalized states which span a large fraction of the gas.

\subsection{Role of interactions in delocalization}

\begin{figure}[t]
\includegraphics[width=\columnwidth]{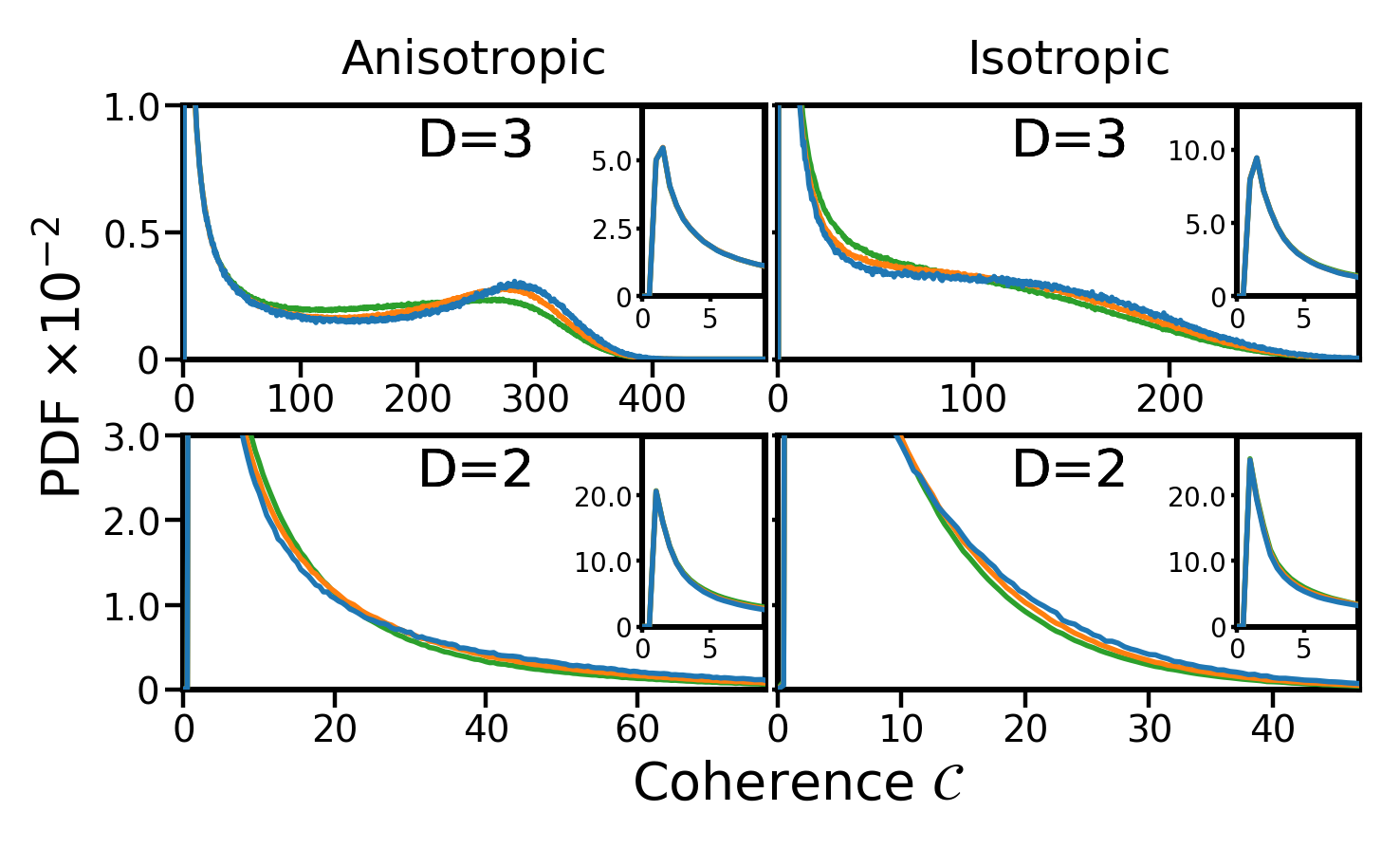}\\
\includegraphics[width=\columnwidth]{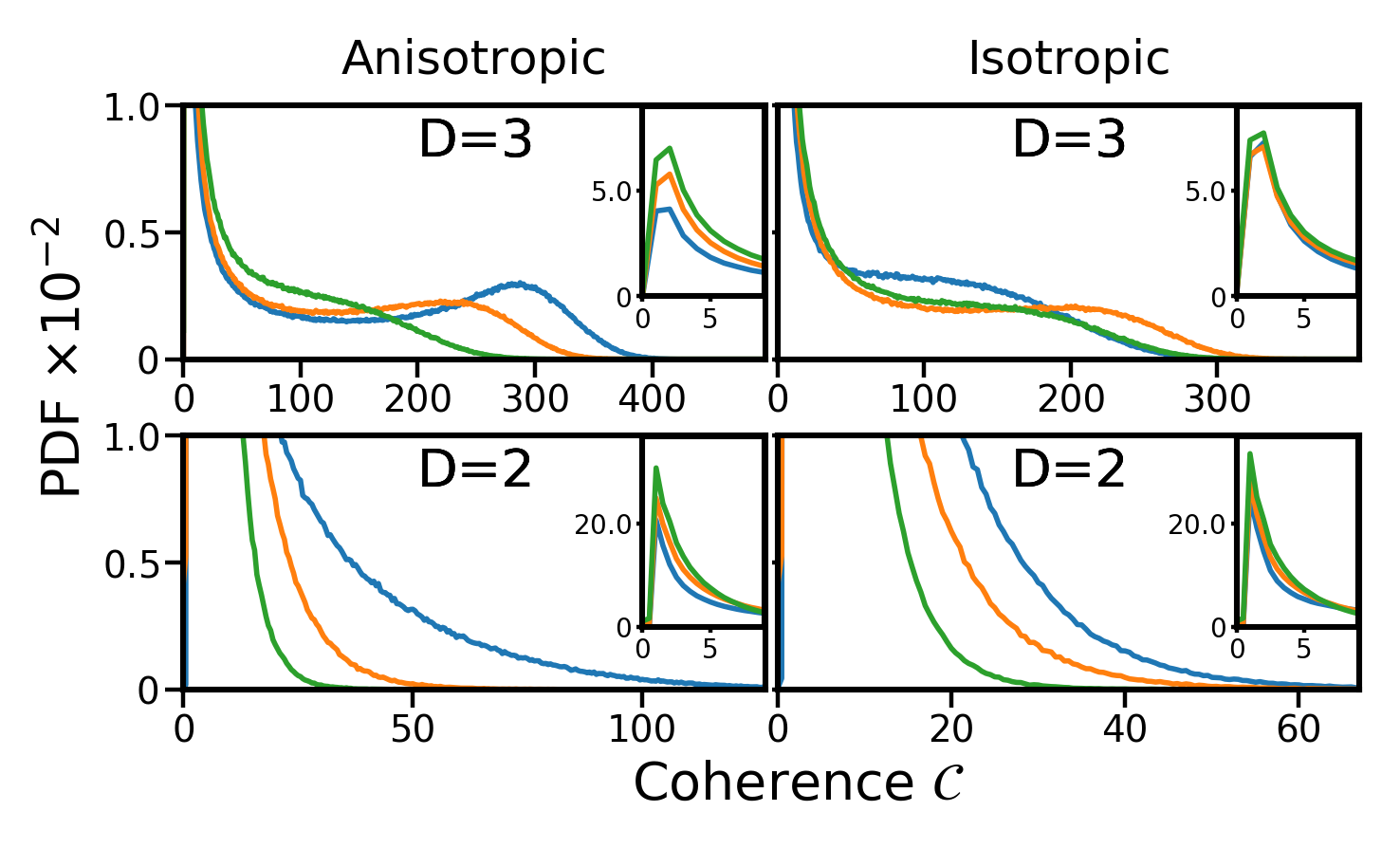}\\
\includegraphics[width=\columnwidth]{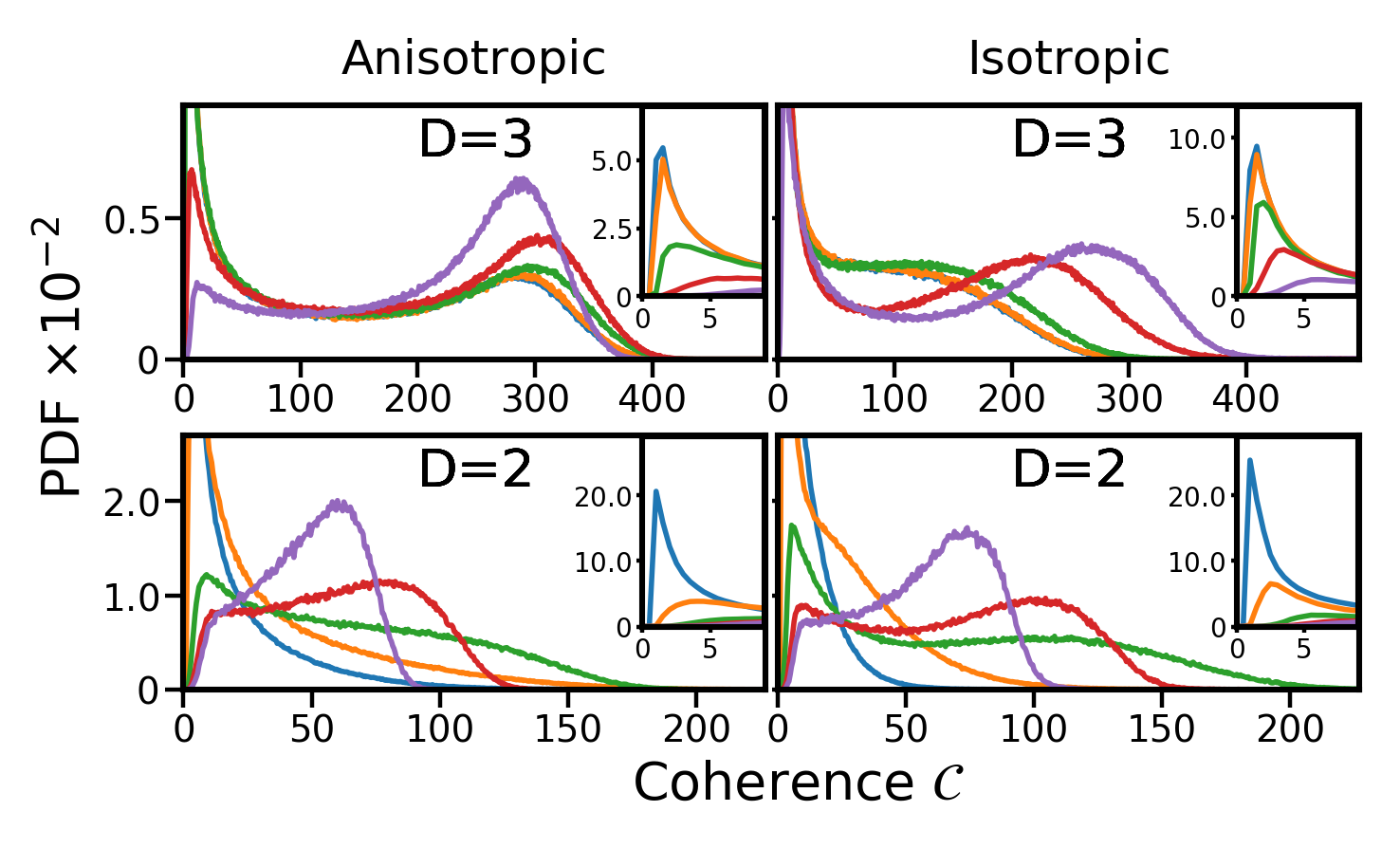}
\caption{Coherence distributions for the $\alpha=3$ interaction with disorder (top), interaction cutoffs (middle), and blockade (bottom). The insets show a closeup of the small coherence region.
Top: four different disorder strengths $V$: 0 (blue), 0.5 (orange), 1.0 (green), all in units of $E_\text{ref}(D)$. The coherence binwidth is 0.5. The insets show a closeup of the small coherence region.
Middle: four different cutoff values $W_c$: 0 (blue), 0.05 (orange), 0.1 (green), all in units of $E_\text{ref}(D)$.  The coherence binwidth is 1. 
Bottom: five different blockade radii ($R_B$): $0$ (blue), 0.25 (orange), 0.5 (green), and 0.75 (red), and 1.0 (purple), all in units of $a_\text{ref}(D)$. The coherence binwidth is 0.5. To compare similar energy scales between two and three-dimensional systems, since the blockade radius is the same in both cases, we changed the density of the two-dimensional system so that $E_\text{ref}(D=2) = E_\text{ref}(D=3)$.}

\label{fig:cutoffs}

\end{figure}

The nature and properties of these delocalized states -- their asymmetry with respect to positive and negative energies, their differences in two and three dimensions, their large sizes -- are linked rather subtly to the matrix elements in the Hamiltonian. In this section we examine three different modifications to these matrix elements in order to test the stability of these states and to reveal aspects of their structure. 

Static energy fluctuations and decoherent processes are known to lead to localization in many systems. A sophisticated study of these effects requires a full inclusion of this physics into the evolution of the density matrix, which is beyond the scope of this article. However, we can perform a proof of principle test of the robustness of these delocalized states by modifying the Hamiltonian to include diagonal disorder, i.e.\ by adding uniformly distributed random values between $\pm V$ to the on-site energies $\varepsilon_n$.  As seen in Fig.~\ref{fig:cutoffs} (top), even at disorder strengths $V$ on the order of the energy unit $E_\text{ref}(D)$ we do not observe any significant changes in the distributions. In the insets we see in particular that the decoupled strongly-interacting cluster distribution is completely unaffected by this disorder. 

As a second check of robustness, we apply a cutoff to the interactions between atoms, setting all matrix elements smaller than a given value $W_c$ to zero. 
Although this model does not correspond to a specific physical mechanism, it tests if the stability of these delocalized states is contingent on very fragile long-range networks bound by weak interactions.  As shown in Fig.~\ref{fig:cutoffs} (middle), we see that the delocalized states are robust against the removal of even quite large interactions, further confirming their stability and also demonstrating that these states are based on extended networks of strongly interacting atoms. We observe that the number of very  small clusters increases with increasing cutoff strength, as evinced by the insets.
We see that the 3D case is less sensitive to the cut-off than the 2D case, and  attribute this to the fact that in 3D each atom can have more neighbors.
 One curious note is that the delocalized states are more stable against this loss in the isotropic case than in the anisotropic, and indeed in three-dimensions the largest coherence lengths even grow when these interactions are clipped. 
In fact, the coherence distributions for the anisotropic case and the isotropic case become nearly equal for the two values of $W_c$  shown in Fig.~\ref{fig:cutoffs}(bottom) although the delocalization distibution in the unperturbed case are quite different with much larger values in the anisotropic case. 
This indicates that in the anisotropic case small interactions are more important than in the isotropic case.

The final interaction modification is the removal of large interactions in the gas, which can be realized experimentally via the Rydberg blockade.  By imposing the blockade at various strengths we prevent the formation of small clusters of strongly interacting atoms on the scale of $R_B$, and hence remove the high-energy tails of the distributions. As seen in the distributions in Fig.~\ref{fig:cutoffs} (bottom) the increasing Rydberg blockade radius manifests itself in a shifting of population from highly localized states into delocalized arrangements. Eventually, and particularly in the two-dimensional system, the Rydberg blockade eliminates all small clusters, as previously seen in Ref. \cite{ourPRL}.

\begin{figure*}[t]
\includegraphics[width=1.\columnwidth]{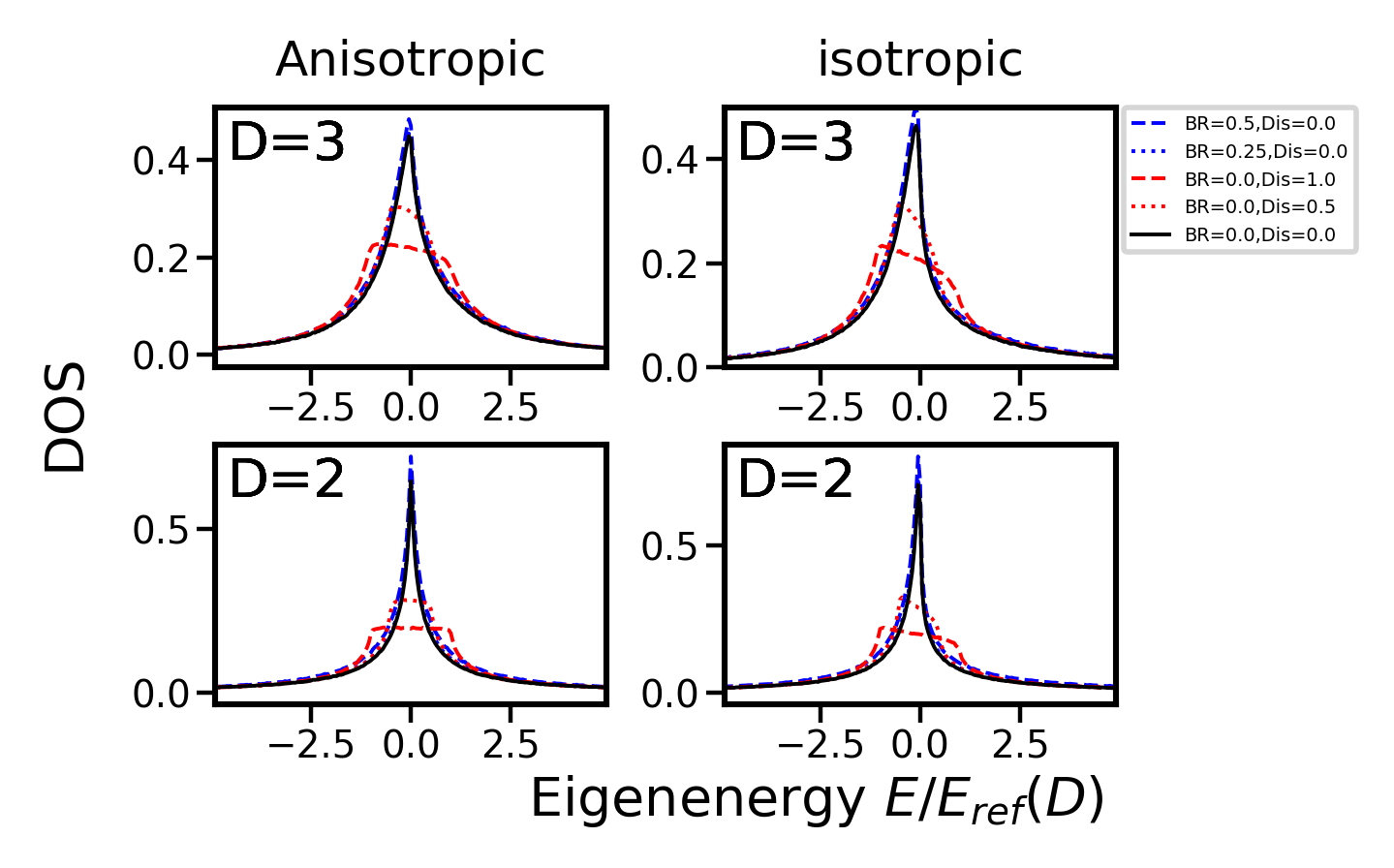}
\includegraphics[width=1.\columnwidth]{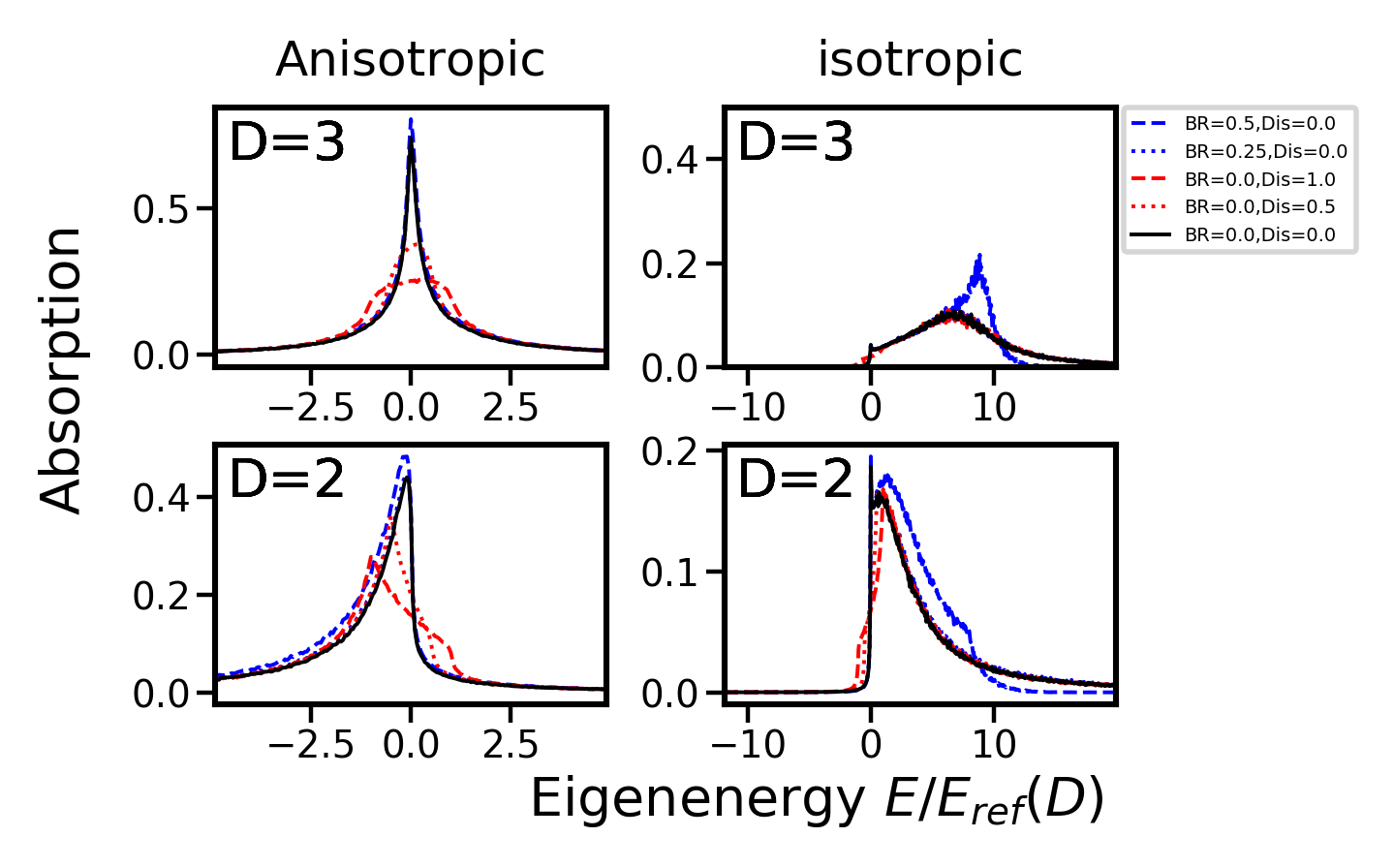}

\caption{Absorption as a function of blockade radius and disorder. The black curve gives the result with $R_B = 0$ and no disorder; red curves have $R_B = 0$ and disorder added with $V=0.5$ $E_\text{ref}(D)$ (dotted)  or $V=1$ $E_\text{ref}(D)$ (dashed). The blue curves have no disorder and $R_B=0.5$ $a_\text{ref}(D)$ (dotted) or $R_B = 0.75$  $a_\text{ref}(D)$ (dashed). }

\label{fig:absorp}

\end{figure*}

\section{Density of states and Absorption spectra}
\label{sec:dos}
It is instructive to compare the energy and coherence resolved plots of  Fig.~\ref{fig:energy} with the density of states (DoS) and the corresponding microwave absorption spectra.
The DoS and absorption spectra are shown in the left and right panels of Fig.~\ref{fig:absorp}, respectively.
The results for the case with no disorder and no Rydberg blockade are shown in black in all panels, while various disorder and blockade strengths are shown in the red and blue curves, respectively.

We first study the DoS. In all four cases it is very similar: it is roughly symmetric and does not reflect the strong asymmetry observed in Fig.~\ref{fig:energy}. 
Note that the DoS is the marginal distribution of the probability density of Fig.~\ref{fig:energy} when one sums over all coherence values.
This highlights the usefulness of the representation of Fig.~\ref{fig:energy}.

The absorption spectra exhibits large differences between all four cases. Note the enlarged energy range of the isotropic case compared to the anisotropic case. While the 3D anisotropic spectrum is roughly symmetric, all other spectra have a strong symmetry. 
The anisotropic 2D case shows strong absorption at negative energies and only very small contributions at small energies.
For the isotropic  cases there is essentially no absorption at negative energies, and in particular the absorption spectrum for the isotropic $D=3$ case bears very little resemblance to its the respective DoS. 
We have found that in all cases delocalized states contribute to the absorption.
In section \ref{sec:analyt} we will use a very simple model to get a basic understanding of these observations.

We now also briefly discuss how disorder and blockade affect the DoS and the absorption spectra.
As expected, the blockade strongly removes contributions in the far wings of DoS and absorption.
For the energy region shown in the DoS this effect is barely visible.
For absorption one sees for the isotropic case even the cutoff (for large blockade radius) at around $E\approx 10$.
Disorder leads to a broadening of the peak at zero energy.

\section{Power law dependence}
\label{sec:power}
\begin{figure*}[t]

\includegraphics[width=\textwidth]{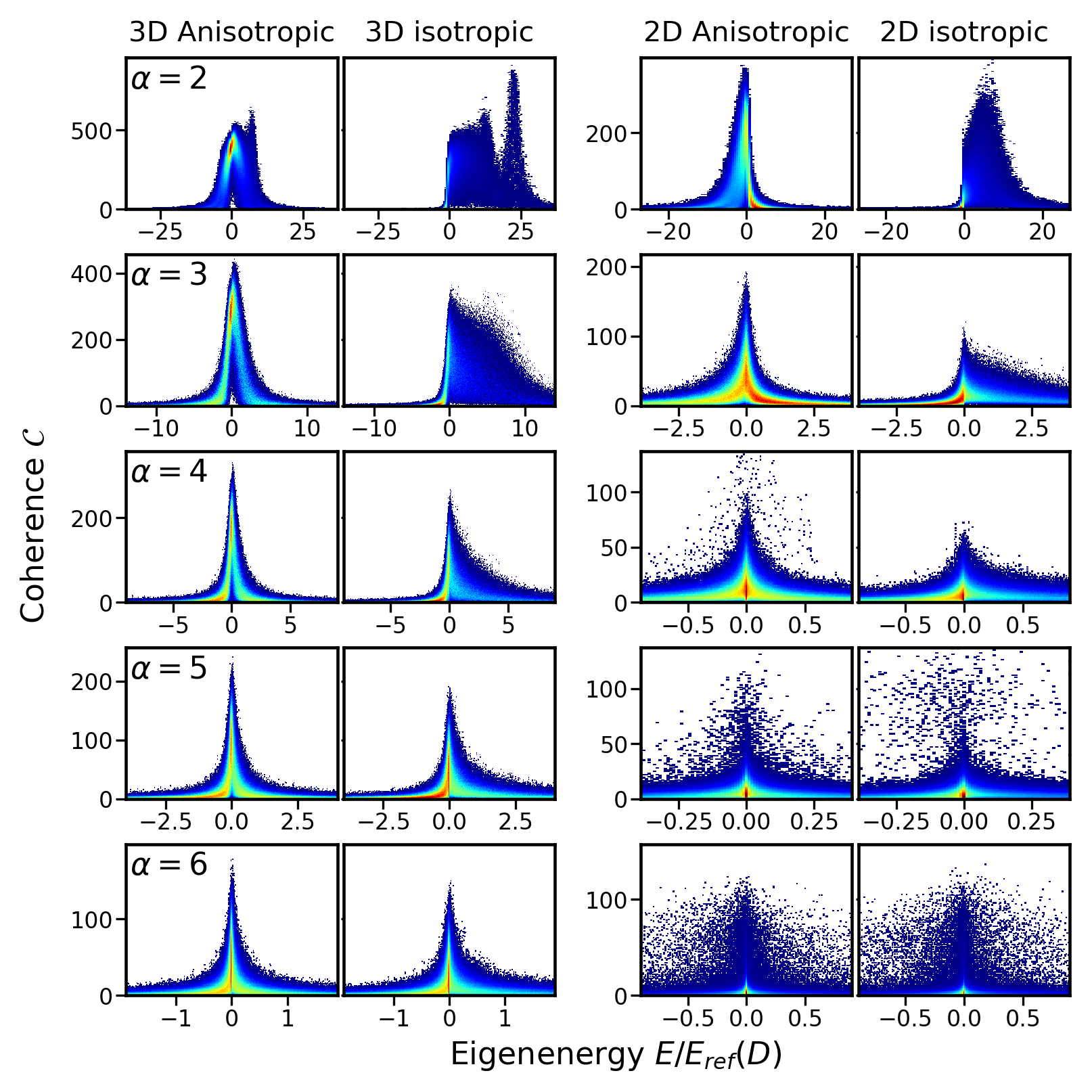}

\caption{Distribution of coherence and eigenenergies for various power-law interactions, in 2D and 3D and with/without anisotropy. The coherence binwidth is 1 and energy binwidth is 0.01.  }

\label{fig:powerlaws}

\end{figure*}

It is instructive to consider also other power law exponents to study the role of the interaction strength and effective range on the delocalization and to better understand the behaviour of the $\alpha=3$ power law. We calculate eigenstates for a variety of power laws, $\alpha = 2,3,\dots 6$, to obtain the coherence and eigenenergy distributions shown in Fig.~ \ref{fig:powerlaws}. Clearly, the coherence lengths and amount of asymmetry increase with the range of the interaction. This asymmetry continues to be strongest in the isotropic case, and particularly for the $\alpha=2$ results is manifested by a nearly instantaneous transition from totally localized dimer states as $E\to 0^-$ to potentialy highly delocalized states as $E\to0^+$. This observation, borne out of the coherence distribution shown here, is supported by a close inspection of many wave functions for many realizations.  We observe an unusual peak structure in the 3D $\alpha=2$ results, which is even more dominant in the $\alpha=1$ case (not shown). While these power-laws seem to be difficult to realize in a Rydberg setup, they might be relevant in other randomly arranged  nanosystems.

\section{Analytical considerations}
\label{sec:analyt}
Some basic features of the observed differences between the isotropic and anisotropic cases can be understood by considering a very simple model of two dimers that weakly interact.
\subsection{Single dimer}
Let us first consider a single dimer.
In the basis of localized excitations $\ket{\up \down}$, $\ket{\down \up}$ the  Hamiltonian can be written as
\be
H=\begin{pmatrix} 0 & V\\ V & 0 \end{pmatrix}
\ee
The eigenstates are $\ket{\psi_\pm}=\frac{1}{\sqrt{2}} (\ket{\up\down} \pm \ket{\down\up}$ with eigenenergies $E_\pm=\pm V$.
The coherence of both eigenstates is $\mathcal{C}_\pm =1$, but the absorption strengths differ greatly: $\mathcal{A}_+= 2$ and $\mathcal{A}_- =0$.
From these results we see that for isotropic interactions (where all interactions are positive) the system's density of states is symmetric around zero and the coherence is the same for positive and negative energies. 
In contrast, only the states with positive energy absorb, producing a strong asymmetry.
This is very different in the case of the anisotropic interaction. Here, depending on the relative orientation of the two atoms with respect to the magnetic field the interaction, $V\sim 1- 3 \cos^2 \theta$, can be positive or negative. 
Therefore also the absorption spectrum becomes more symmetric, because for negative dipole-dipole coupling $V$ the absorbing state is at negative energies.
This model explains the regions of large eigenenergies, which is dominated by dimers, in the case of much higher $N$. It is also helpful to understand the following model of coupled dimers.

\subsection{Coupled dimers}
We consider now the case of four atoms.
As for the dimer case above, in the basis of localized excitations the Hamiltonian can be written as
\be
H = \begin{pmatrix}0 & V_{12} & V_{13} & V_{14}\\ V_{12} & 0 & V_{23} &V_{24} \\  V_{13} & V_{23}& 0 & V_{34} \\ V_{14} & V_{24} & V_{34} & 0\end{pmatrix},
\ee
For general arrangements of the atoms one cannot get a simple analytical form of the eigenstates.
Therefore, we will now assume that  $V_{23} = V_{13} = V_{14} = V_{24} \equiv  W$. 
 In  matrix form this reads
\be
H = \begin{pmatrix}0 & V_A & W & W\\ V_A & 0 & W &W \\  W & W& 0 & V_B \\ W & W & V_B & 0\end{pmatrix},
\ee
where $V_A\equiv V_{12}$ and $V_B\equiv V_{34}$.
This Hamiltonian corresponds in the isotropic case to the situation where atoms 1 and 2 are close, atoms 3 and 4 are close, and the two pairs of atoms are so far apart that interactions between the two dimers are approximately equal. 
Clearly, $W\ll V_A$ and $W\ll V_B$.
 With this simple model we are able to draw some general conclusions. 
 
Let us first look at the case $V_A=V_B\equiv V$, which gives analytical results amenable to analysis. 
 The four eigenvalues are $-V$, $-V$, and $V \pm 2W$. Note that the coupling is asymmetric: the  dimer energies $-V$ from the two-atom model are unshifted, while the coupling modifies the two energies $+V$. 
The eigenstates corresponding to the eigenvalue $-V$ are both antisymmetric dimer states (coherence $\mathcal{C}=1$) with eigenvectors $2^{-1/2}(-1,1,0,0)$ or $2^{-1/2}(0,0,-1,1)$.
This implies that there is no absorption strength associated with these states.
The eigenstate corresponding to the energy $V - 2W$ is $2^{-1}(-1,-1,1,1)$ and also does not have absorption strength. 
On the other hand, the state at energy $V + 2W$ is $2^{-1}(1,1,1,1)$ and has an absorption stength wich is four times that of a single atom.
Note that these two states are fully delocalized with coherence $\mathcal{C}=3$.
For the isotropic case these results imply that, as in the dimer case, there is absorption only for positive energies, which corresponds to the numerical observation of Fig.~\ref{fig:absorp}.
Furthermore, at positive energies the coherence is much larger than at negative energies, which is also in general agreement with Fig.~\ref{fig:energy}.
For the anisotropic case the situation becomes more complicated, since $V$ and $W$ can both be positive and negative.
When $V$ is positive the delocalized absorbing state is on the positive side (we consider $V>W$).
For negative $V$ the delocalized absorbing state is at negative energies. 
This leads to results that are rather symmetric with respect to the zero of energy. 

Lets now consider the case $V_A \ne V_B$. 
The resulting eigenenergies and eigenstates are given in Table \ref{tab:coupledDimer}.
One sees again that there are two dimer eigenstates ($\ell=1,2$) with coherence $\mathcal{C}=1$ that do not absorb.
For the other two states the situation is more complicated.
In the isotropic case $V_{\rm mean}$ is positive and thus the states $\ell=3,4$ are located at positive energies. 
When $V_A$ and $V_B$ are similar, then $\tilde{V}_{\rm diff}$ is small and $\tilde{S} \approx 1$. Both states are delocalized and the state 4 has a large absorption strength while state 3 has a small absorption strength.
When $\tilde{V}_{\rm diff}$ is larger than 1, then the coherence length   approaches -- for increasing $\tilde{V}_{\rm diff}$ -- dimer-like coherence. 
That means that for isotropic interactions the absorbing states are at positive energies and can have coherences ranging from 1 to the maximal coherence.
For the anisotropic case a variety of combinations of the signs and magnitudes of $V_A$, $V_B$ and $W$ is possible.
Now, if $V_A \approx V_B$ it can happen that $V_{\rm mean}\approx 0$. 
Then one sees that the delocalized and absorbing states are situated around the zero of energy. 
The discussion presented above provides a basic understanding of the observations made in Fig.~\ref{fig:energy} and Fig.~\ref{fig:absorp}.
A more detailed discussion (e.g.~by integrating over angular distributions) would go beyond the aim of the present work, but would likely be a fruitful avenue to pursue.

\begin{table*}
\begin{tabular}{c|c|c|c|c}
 $\ell$& $E_\ell$ & $\psi_\ell$ & $\mathcal{C}_\ell$ & $\mathcal{A}_\ell$ \\
\hline
1   &   $-V_A$ & $(1,-1,0,0)$ &  1 & 0 \\
2   &   $-V_B $ & $(0,0,1,-1)$ & 1 & 0 \\
3 &     $ V_{\rm mean}- \frac{S}{2} $ &  $(\tilde{V}_{\rm diff}-\tilde{S},  \tilde{V}_{\rm diff}-\tilde{S},1,1)$  & $1+\frac{8W}{S}$ & $2 - \frac{8W}{S}$ \\
4  &    $V_{\rm mean}+ \frac{S}{2}$ & $(\tilde{V}_{\rm diff}+\tilde{S},  \tilde{V}_{\rm diff}+\tilde{S},1,1)$  & $1+\frac{8W}{S}$ & $2 + \frac{8W}{S}$
\end{tabular}
\caption{\label{tab:coupledDimer} Properties of the coupled dimer model discussed in the text. We have defined:
$V_{\rm mean}= (V_A+V_B)/2$, $V_{\rm diff}= V_A - V_B$, $S=\sqrt{V_{\rm diff}^2+ 16 W^2 }$. 
}
\end{table*}

\section{Conclusions}
\label{sec:conc}
Through our study of the eigenstates of an interacting random Rydberg gas, we have seen that the surprisingly high amount of delocalization in these states first observed in Ref.~\cite{ourPRL} persists in both 3D and 2D gases. These delocalized states are very robust against a range of disorder or decoherence processes. The Rydberg blockade, by creating a more regular arrangement array of background $\down$ atoms, leads to even larger degrees of delocalization. These delocalized states persist for the various power-law interactions studied here. 
They are less pronounced when the interaction becomes isotropic.
We have used a simple model of coupled dimers to provide basic insight into the strong differences between the isotropic and anisotropic case. In the future, it would be useful to extend these models to try to capture some of the distinctions between two and three-dimensional configurations. 

The Hamiltonians underlying the present study fall into the class of so called ``Euclidean random matrices'' \cite{mezard1999spectra,Akulin2014}. 
For such matrices powerfull analytical tools have been developed \cite{Akulin2014}, in particular to describe the density of states. 
It would be interesting to compare the numerical findings of the specific systems of the present work with simple analytical considerations going beyond the model of Sec.~\ref{sec:analyt}. In particular, an extension of the analytic results for eigenvalues given by these studies of Euclidean random matrices to the eigenstates could provide useful insight. 

We have seen that density of states and absorption are not sufficient to obtain direct information about the delocalization properties of the eigenstates.  As demonstrated in Ref.~\cite{ourPRL}, one can use strong microwave pulses of well-determined durations and Rabi frequencies to selectively populate delocalized states and probe them directly. The existence of such delocalized states has implications for non-linear spectroscopy of many-body states in randomly arranged systems since the localization of many-body states will be determined by the extent of the underlying one-exciton domains.

\ack{

We acknowledge funding from the DFG: grant EI 872/4-1 through the Priority Programme SPP 1929 (GiRyd). AE acknowledges support from the DFG via a Heisenberg fellowship (Grant No EI 872/5-1). MTE acknowledges support from an Alexander von Humboldt Stiftung postdoctoral fellowship.  }

\section*{Appendix I: Comparison of coherence and participation ratio}
\label{appendix}
\begin{figure}[h]

\includegraphics[width=\columnwidth]{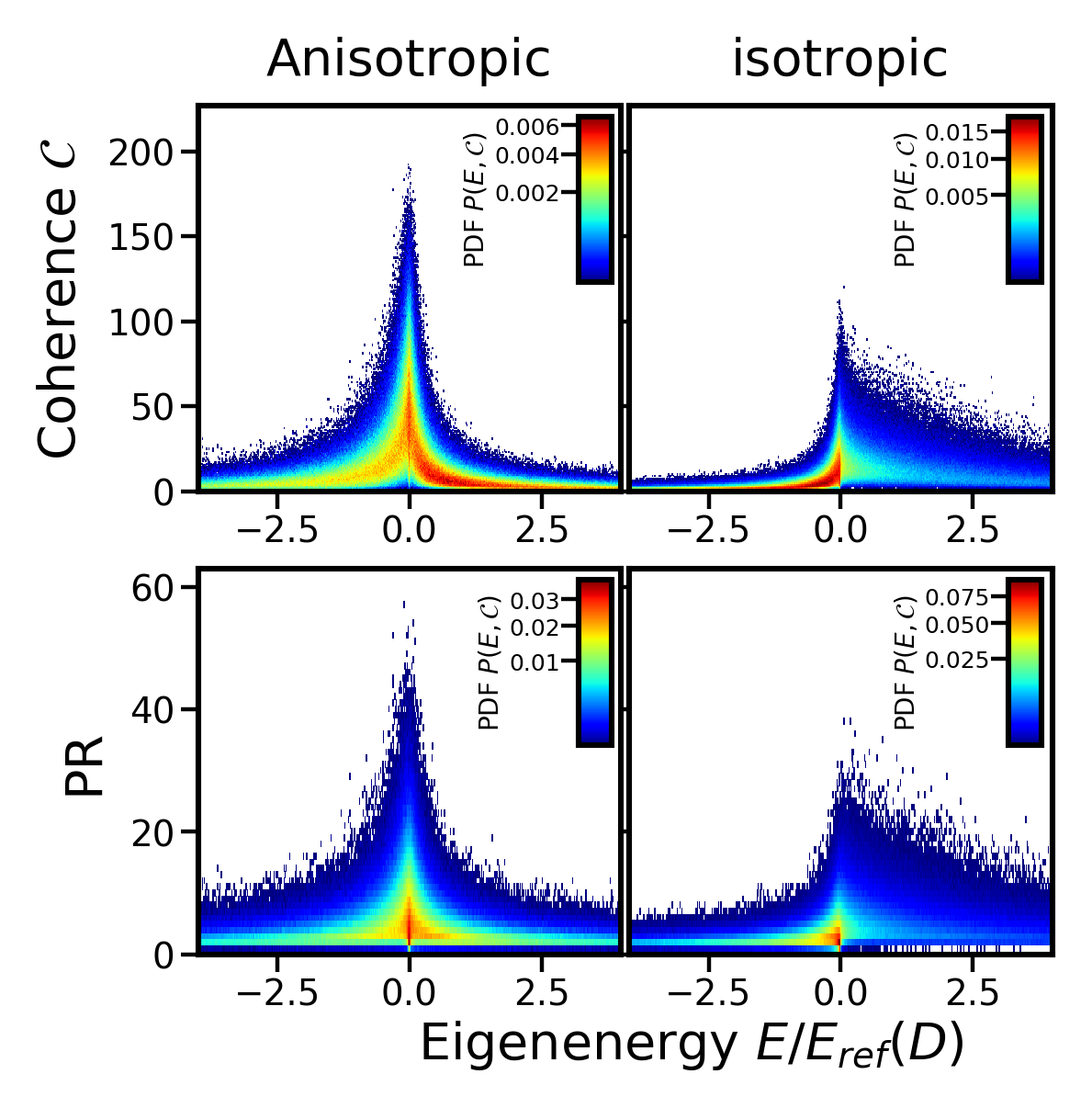}

\caption{Distribution of coherence values and eigenenergies (top) and PR and eigenenergies (bottom) for the two-dimensional gas. The coherence and PR binwidth is 0.5 and the energy binwidth is 0.01.   }

\label{fig:iprappendix}

\end{figure}

In this appendix we revisit some of the results shown in the main text, but now using PR rather than coherence as the measure of delocalization. 
Fig.~\ref{fig:iprappendix} compares the distributions of coherence values with the distributions of PRs for the two-dimensional gas. The distributions are rather similar in overall shape and structure except for an overall scale factor. However, closer inspection reveals that the PR does not exhibit the strong asymmetry in the localized states close to zero eigenenergy that is so evident in the coherence; this furthermore limits its utility in differentiating the isotropic and anisotropic coherence distributions.

\begin{figure}

\includegraphics[width=\columnwidth]{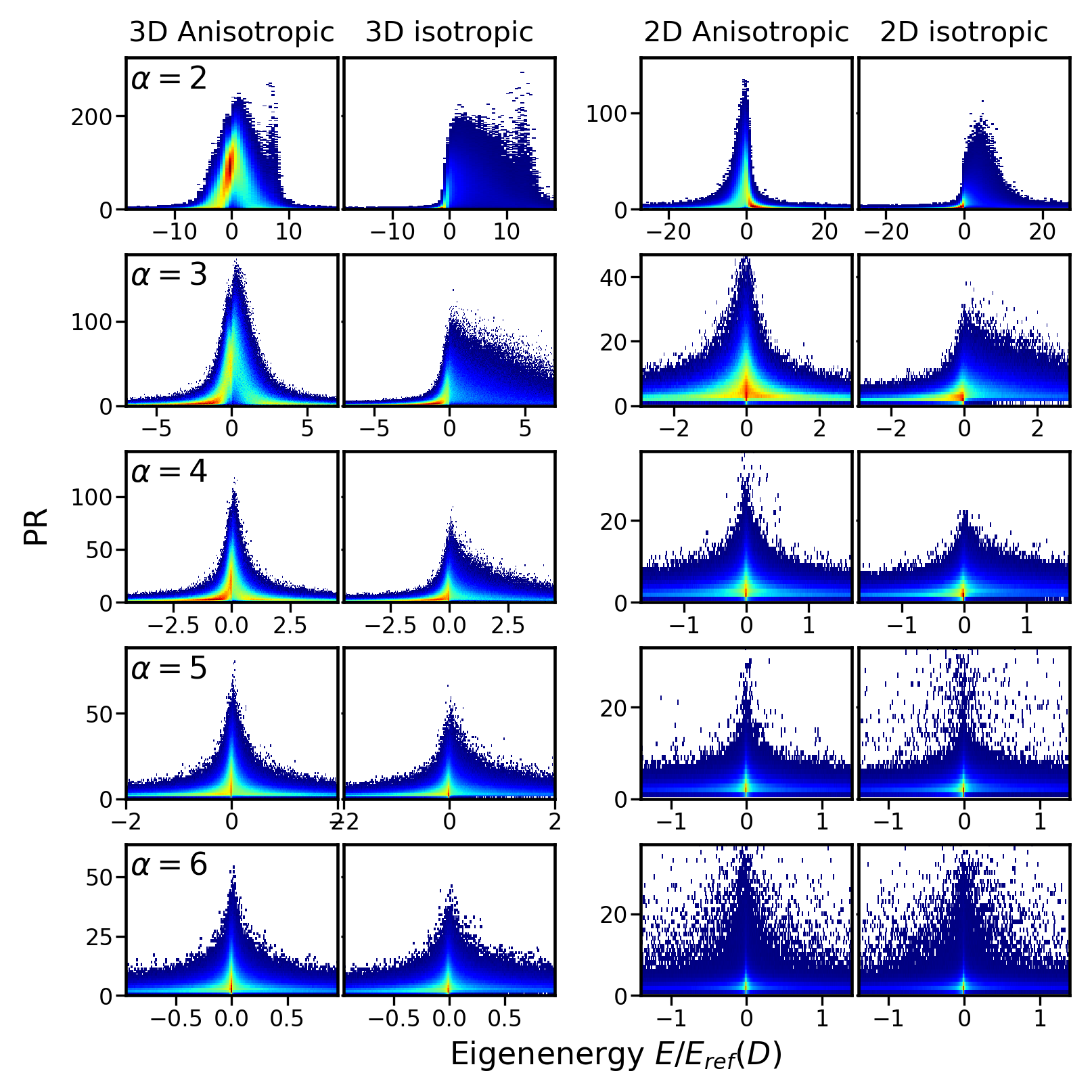}

\caption{PR and energy distributions for the the various power-law interactions. The PR binwidth is 0.5 and the energy binwidth is 0.01. }

\label{fig:IPRenergy}

\end{figure}

This is true for all power law interactions, as seen in Fig.~\ref{fig:IPRenergy}, although in the 3D case the distribution is still markedly asymmetric. For this reason, although for most features the PR and coherence provide equivalent information about delocalization, coherence is a slightly more sensitive measure especially in two dimensions.

\section*{References}

\begin{thebibliography}{10}
\expandafter\ifx\csname url\endcsname\relax
  \def\url#1{{\tt #1}}\fi
\expandafter\ifx\csname urlprefix\endcsname\relax\def\urlprefix{URL }\fi
\providecommand{\eprint}[2][]{\url{#2}}

\bibitem{ourPRL}
Abumwis G, Eiles M~T and Eisfeld A 2019 {\em arXiv:1909.12705\/}

\bibitem{SaEiVa13_21_}
Saikin S~K, Eisfeld A, Valleau S and Aspuru-Guzik A 2013 {\em Nanophotonics\/}
  {\bf 2} 21

\bibitem{AmVaGr00__}
van Amerongen H, Valkunas L and van Grondelle R 2000 {\em {Photosynthetic
  Excitons}\/} (World Scientific, Singapore)

\bibitem{yan2013}
Yan B, Moses S~A, Gadway B, Covey J~P, Hazzard K~R, Rey A~M, Jin D~S and Ye J
  2013 {\em Nature\/} {\bf 501} 521

\bibitem{BaLaRa15_113002_}
Barredo D, Labuhn H, Ravets S, Lahaye T, Browaeys A and Adams C~S 2015 {\em
  Phys. Rev. Lett.\/} {\bf 114} 113002

\bibitem{DaRiLi12_193201_}
Dai X, Richter M, Li H, Bristow A~D, Falvo C, Mukamel S and Cundiff S~T 2012
  {\em Phys. Rev. Lett.\/} {\bf 108}(19) 193201

\bibitem{BrBiSt15_053412_}
Bruder L, Binz M and Stienkemeier F 2015 {\em Phys. Rev. A\/} {\bf 92} 053412

\bibitem{BrEiBa19_-_}
Bruder L, Eisfeld A, Bangert U, Binz M, Jakob M, Uhl D, Schulz-Weiling M, Grant
  E~R and Stienkemeier F 2019 {\em Phys. Chem. Chem. Phys.\/}  2276--2282

\bibitem{Cundiff2016}
Feng G, Cundiff S~T and Hebin L 2016 {\em Opt. Lett.\/} {\bf 41} 2954--2957

\bibitem{yu2019long}
Yu S, Titze M, Zhu Y, Liu X and Li H 2019 {\em Optics Express\/} {\bf 27}
  28891--28901

\bibitem{Mu16_041102_}
Mukamel S 2016 {\em The Journal of Chemical Physics\/} {\bf 145} 041102

\bibitem{LiBrSt17_052509_}
Li Z~Z, Bruder L, Stienkemeier F and Eisfeld A 2017 {\em Phys. Rev. A\/} {\bf
  95} 052509

\bibitem{Pillet1998}
Mourachko I, Comparat D, de~Tomasi F, Fioretti A, Nosbaum P, Akulin V~M and
  Pillet P 1998 {\em Phys. Rev. Lett.\/} {\bf 80}(2) 253--256

\bibitem{GallagherFroze1998}
Anderson W~R, Veale J~R and Gallagher T~F 1998 {\em Phys. Rev. Lett.\/} {\bf
  80}(2) 249--252

\bibitem{Mourachko2004}
Mourachko I, Li W and Gallagher T~F 2004 {\em Phys. Rev. A\/} {\bf 70}(3)
  031401

\bibitem{scholak2014spectral}
Scholak T, Wellens T and Buchleitner A 2014 {\em Physical Review A\/} {\bf 90}
  063415

\bibitem{Pilleta}
Pillet P, Comparat D, Muldrich M, Vogt T, Zahzam N, Akulin V~M, Gallagher T~F,
  Li W, Tanner P, Noel M~W and Mourachko I 2005 Coherence and decoherence in
  rydberg gases {\em Decoherence, Entanglement and Information Protection in
  Complex Quantum Systems\/} (Springer-Verlag) pp 411--436

\bibitem{Singer2005}
Singer K, Stanojevic J, Weidem{\"u}ller M and C{\^{o}}t{\'{e}} R 2005 {\em
  Journal of Physics B: Atomic, Molecular and Optical Physics\/} {\bf 38}
  S295--S307

\bibitem{Weber_2017}
Weber S, Tresp C, Menke H, Urvoy A, Firstenberg O, B{\"u}chler H~P and
  Hofferberth S 2017 {\em Journal of Physics B: Atomic, Molecular and Optical
  Physics\/} {\bf 50} 133001

\bibitem{MartinDDSpec}
Afrousheh K, Bohlouli-Zanjani P, Vagale D, Mugford A, Fedorov M and Martin
  J~D~D 2004 {\em Phys. Rev. Lett.\/} {\bf 93}(23) 233001

\bibitem{Park2011}
Park H, Tanner P~J, Claessens B~J, Shuman E~S and Gallagher T~F 2011 {\em
  Physical Review A\/} {\bf 84}

\bibitem{Robicheaux2004}
Robicheaux F, Hern{\'{a}}ndez J~V, Top{\c{c}}u T and Noordam L~D 2004 {\em
  Physical Review A\/} {\bf 70}

\bibitem{chomaz2015emergence}
Chomaz L, Corman L, Bienaim{\'e} T, Desbuquois R, Weitenberg C, Nascimb{\`e}ne
  S, Beugnon J and Dalibard J 2015 {\em Nature communications\/} {\bf 6} 6162

\bibitem{MPaEi13_044302_}
M\"{u}ller M, Paulheim A, Eisfeld A and Sokolowski M 2013 {\em J. Chem.
  Phys.\/} {\bf 139} 044302

\bibitem{MPaMa13_064703_}
M\"{u}ller M, Paulheim A, Marquardt C and Sokolowski M 2013 {\em J. Chem.
  Phys.\/} {\bf 138} 064703

\bibitem{PaTaCl11_22704_}
Park H, Tanner P~J, Claessens B~J, Shuman E~S and Gallagher T~F 2011 {\em Phys.
  Rev. A\/} {\bf 84} 022704

\bibitem{low2012experimental}
L{\"o}w R, Weimer H, Nipper J, Balewski J~B, Butscher B, B{\"u}chler H~P and
  Pfau T 2012 {\em Journal of Physics B: Atomic, Molecular and Optical
  Physics\/} {\bf 45} 113001

\bibitem{saffmanJPB}
Saffman M 2016 {\em Journal of Physics B: At. Mol. Opt. Phys.\/} {\bf 49}
  202001

\bibitem{Saffman}
Saffman M, Walker T~G and M\"{o}lmer K 2010 {\em Rev. Mod. Phys.\/} {\bf 82}
  2313

\bibitem{LiTaJa06_27_}
Li W, Tanner J~P, Jamil Y and Gallagher F~T 2006 {\em Eur. Phys. J. D\/} {\bf
  40} 27--35

\bibitem{Li2006}
Li W, Tanner P~J, Jamil Y and Gallagher T~F 2006 {\em The European Physical
  Journal D\/} {\bf 40} 27--35

\bibitem{Plasma1}
Li W, Noel M~W, Robinson M~P, Tanner P~J, Gallagher T~F, Comparat D,
  Laburthe~Tolra B, Vanhaecke N, Vogt T, Zahzam N, Pillet P and Tate D~A 2004
  {\em Phys. Rev. A\/} {\bf 70}(4) 042713

\bibitem{Plasma2}
Li W, Tanner P~J and Gallagher T~F 2005 {\em Phys. Rev. Lett.\/} {\bf 94}(17)
  173001

\bibitem{AtEiRo08_045030_}
Ates C, Eisfeld A and Rost J~M 2008 {\em New J. Phys.\/} {\bf 10} 045030

\bibitem{WEiRo11_153002_}
W\"uster S, Eisfeld A and Rost J~M 2011 {\em Phys. Rev. Lett.\/} {\bf 106}
  153002

\bibitem{BaCrPl14_140401_}
Baumgratz T, Cramer M and Plenio M~B 2014 {\em Phys. Rev. Lett.\/} {\bf
  113}(14) 140401

\bibitem{Lukin2001}
Lukin M~D, Fleischhauer M, Cote R, Duan L~M, Jaksch D, Cirac J~I and Zoller P
  2001 {\em Physical Review Letters\/} {\bf 87}

\bibitem{tong2004local}
Tong D, Farooqi S, Stanojevic J, Krishnan S, Zhang Y, C{\^o}t{\'e} R, Eyler E
  and Gould P 2004 {\em Physical Review Letters\/} {\bf 93} 063001

\bibitem{gaetan2009observation}
Ga{\"e}tan A, Miroshnychenko Y, Wilk T, Chotia A, Viteau M, Comparat D, Pillet
  P, Browaeys A and Grangier P 2009 {\em Nature Physics\/} {\bf 5} 115

\bibitem{urban2009observation}
Urban E, Johnson T~A, Henage T, Isenhower L, Yavuz D, Walker T and Saffman M
  2009 {\em Nature Physics\/} {\bf 5} 110

\bibitem{ates2007}
Ates C, Pohl T, Pattard T and Rost J~M 2007 {\em Physical review letters\/}
  {\bf 98} 023002

\bibitem{pohl2010}
Pohl T, Demler E and Lukin M~D 2010 {\em Physical review letters\/} {\bf 104}
  043002

\bibitem{schauss2012}
Schau{\ss} P, Cheneau M, Endres M, Fukuhara T, Hild S, Omran A, Pohl T, Gross
  C, Kuhr S and Bloch I 2012 {\em Nature\/} {\bf 491} 87

\bibitem{mezard1999spectra}
M{\'e}zard M, Parisi G and Zee A 1999 {\em Nuclear Physics B\/} {\bf 559}
  689--701

\bibitem{Akulin2014}
Akulin V~M 2014 {\em Dynamics of Complex Quantum Systems\/} (Springer
  Netherlands)

\end{thebibliography}

\providecommand{\newblock}{}

\end{document}